\documentclass[letter]{aa} % for the letters 
\usepackage{graphicx}
\usepackage{txfonts}
\usepackage{lscape}
\usepackage{mathrsfs}
\usepackage{nicefrac}
\begin{document}
\title{Stellar mass estimates in early-type galaxies from lensing+dynamical and photometric measurements}

\author{C. Grillo\inst{1,2}, R. Gobat\inst{1}, P. Rosati\inst{1}, and M. Lombardi\inst{1,2}}

\offprints{C. Grillo}

\institute{European Southern Observatory, Karl-Schwarzschild-Str. 2, D-85748, Garching bei M\"unchen, Germany\\
  \email{cgrillo@eso.org}
  \and
  Universit\`a degli Studi di Milano, Department of Physics,
  via Celoria 16, I-20133 Milan, Italy\\
  %\email{claudio.grillo@unimi.it}
}

\authorrunning{C.~Grillo et al.}

\titlerunning{Stellar mass estimates in early-type galaxies from lensing+dynamical and photometric measurements}

\date{Received X X, X; accepted Y Y, Y}

% \abstract{}{}{}{}{} 
% 5 {} token are mandatory
 
  \abstract
   {}
   {To compare two different diagnostics for estimating stellar masses in early-type galaxies and to establish their level of reliability. In particular, we consider the well-studied sample of 15 field elliptical galaxies selected from the Sloan Lens ACS (SLACS) Survey ($z = 0.06-0.33$). We examine here the correlation between the stellar mass values, enclosed inside the Einstein radius ($R_{\mathrm{Ein}}$) of each lens, based on analyses of lensing and stellar dynamics combined and based on multiwavelength photometry spectral template fitting.}
   {The lensing+dynamics stellar mass $M_{\mathrm{len+dyn}}^{*}(\le R_{\mathrm{Ein}})$ is obtained from the published SLACS Survey results, assuming a two-component density distribution model and a prior from the fundamental plane on the mass-to-light ratio for the lens galaxies. The photometric stellar mass $M_{\mathrm{phot}}^{*}(\le R_{\mathrm{Ein}})$ is measured by fitting the observed spectral energy distribution of the galaxies (from the SDSS multi-band photometry over $354-913$ nm) with composite stellar population templates, under the assumption that light traces stellar mass.}
   {The two methods are completely independent. They rely on several different assumptions, and so, in principle, both can have significant biases. Based on our sample of massive galaxies ($\mathrm{log}M_{\mathrm{phot}}^{*}(\le R_{\mathrm{Ein}})\simeq[10.3,11.5]$), we find consistency between the values of $M_{\mathrm{len+dyn}}^{*}(\le R_{\mathrm{Ein}})$ and $M_{\mathrm{phot}}^{*}(\le R_{\mathrm{Ein}})$. We obtain a Pearson linear correlation coefficient of 0.94 and a median value of the ratio between the former and the latter mass measurements of $1.1\pm0.1$. This suggests that both methods can separately yield reliable stellar masses of early-type galaxies, and confirms that photometric mass estimates are accurate, as long as optical/near-IR rest frame photometry is available.}
   {}


\maketitle
%
%________________________________________________________________

\section{Introduction}

An estimate of the stellar mass component in galaxies is interesting for several different reasons. In detail, by combining or comparing photometric stellar mass estimates, obtained by spectral energy distribution (SED) fitting methods, with dynamical or lensing measurements, it is possible to study the radial distribution of dark matter (e.g., Ferreras et al. \cite{fer05}; Napolitano et al. \cite{nap05}), to investigate the relationship between stellar and total mass (e.g., Lintott et al. \cite{lin06}; Rettura et al. \cite{ret06}), and to test hierarchical structure formation models (e.g., Nagamine et al. \cite{nag04}; De Lucia et al. \cite{del06}). Interestingly, Treu \& Koopmans (\cite{tre2}) have proved that the stellar mass fraction in elliptical lens galaxies can also be estimated with a joint lensing and dynamical analysis. 

Although it is common to measure stellar masses through these techniques, only a few studies have been performed to check the reliability of each method (e.g., Drory et al. \cite{dro04}). Further investigations are therefore important to probe the consistency of the different techniques.

Throughout this work we assume $H_{0}=70 \mbox{ km s}^{-1} \mbox{ Mpc}^{-1}$, $\Omega_{m}= 0.3$, and $\Omega_{\Lambda} = 0.7$.

\section{The SLACS sample}

In this Letter, we focus on a uniformly selected sample of 15 massive field early-type galaxies taken from the SLACS Survey (for more details on the selection procedure, see Bolton et al. \cite{bol}). Table \ref{Infotable} summarizes the relevant photometric and spectroscopic properties of the galaxy sample.
The lens galaxies have a redshift between 0.06 and 0.33, \emph{HST} F435W and F814W images, \emph{ugriz} magnitudes and stellar velocity dispersions from the SDSS\footnote{http://www.sdss.org/}. They are luminous red galaxies (LRG; Eisenstein et al. \cite{eis}) with properties similar to those of non-lensing early-type galaxies: redshifts, stellar velocity dispersions, stellar populations, and mass density profiles (see Treu et al. \cite{tre1}; Koopmans et al. \cite{koo3}). 

%__________________________________________________________________

\section{Measuring stellar masses}

Here, we describe how the stellar mass of the galaxies of our sample is measured using two independent diagnostics.

\subsection{Lensing+dynamical estimates}

Strong gravitational lensing provides the most accurate estimate of the total (stellar+dark) projected mass of a lens galaxy inside the Einstein radius. It has been shown by Treu \& Koopmans (\cite{tre2}) that by combining lensing measurements with spatially resolved kinematic profiles in elliptical galaxies, the stellar and dark components can be separated precisely. If the velocity dispersion of stars is known only from a single (fiber) aperture, some information on the stellar mass fraction ($f_{*}$) inside $R_{\mathrm{Ein}}$ can still be obtained. This particular analysis has been performed on the SLACS sample by Koopmans et al. (\cite{koo3}). The results are shown in Table \ref{Masstable}. We summarize here the main steps and assumptions:
\begin{itemize}

\item The mass distribution of each lens galaxy is described in terms of a two-component spherical and isotropic model. It consists of a Hernquist density profile, scaled by a stellar mass-to-light ratio $M_{*}/L$, for the stellar component, and a power-law density profile, for the dark matter component ($\rho_{\mathrm{d}}=\rho_{\mathrm{d,0}}\,r^{-\gamma}$).

\item The lensing measurement of the total projected mass enclosed inside the Einstein circle, $M_{\mathrm{len+dyn}}^{\mathrm{tot}}(\le R_{\mathrm{Ein}})$, is used to eliminate $\rho_{\mathrm{d,0}}$. Thus, for any given set $\{M_{*}/L,\gamma\}$, the spherical Jeans equation can be solved to determine the line-of-sight stellar velocity dispersion as a function of radius.

\item A likelihood function is defined by comparing the model predicted with the observed aperture stellar velocity dispersion. A prior on the stellar mass-to-light ratio, based on the fundamental plane (in the $B$ band) and corrected for passive evolution, is considered before marginalizing on the two free parameters of the model ($M_{*}/L$ and $\gamma$).

\item The stellar mass fraction is calculated as the ratio between the maximum likelihood and the maximum allowed values of $M_{*}/L$. This latter quantity is obtained under the assumption that $M_{\mathrm{len+dyn}}^{\mathrm{*}}(\le R_{\mathrm{Ein}})=M_{\mathrm{len+dyn}}^{\mathrm{tot}}(\le R_{\mathrm{Ein}})$.

\end{itemize}
In principle, this method can give values of $f_{*}$ bigger than 1, as can be seen in a few cases in Table \ref{Masstable}. Nevertheless, these values are consistent with 1, so that the previous assumptions are not in doubt. Finally, Koopmans et al. (\cite{koo3}) point out that more realistic values of $M_{*}/L$ should take into account the dependence on the mass of the galaxy.

The stellar mass obtained from the combination of gravitational lensing and stellar dynamics measurements inside the Einstein radius, $M_{\mathrm{len+dyn}}^{*}(\le R_{\mathrm{Ein}})$, is then estimated by multiplying the total mass by the stellar mass fraction. This is shown in the following equation
\begin{equation}
M_{\mathrm{len+dyn}}^{*}(\le R_{\mathrm{Ein}}) = f_{*}\,\times\,M_{\mathrm{len+dyn}}^{\mathrm{tot}}(\le R_{\mathrm{Ein}})\,,
\label{eq:4}
\end{equation}
and the results for the SLACS sample (Koopmans et al. \cite{koo3}) are reported in Table \ref{Masstable}.

\subsection{Photometric estimates}

\begin{figure}
  \centering
  \includegraphics[width=0.49\textwidth]{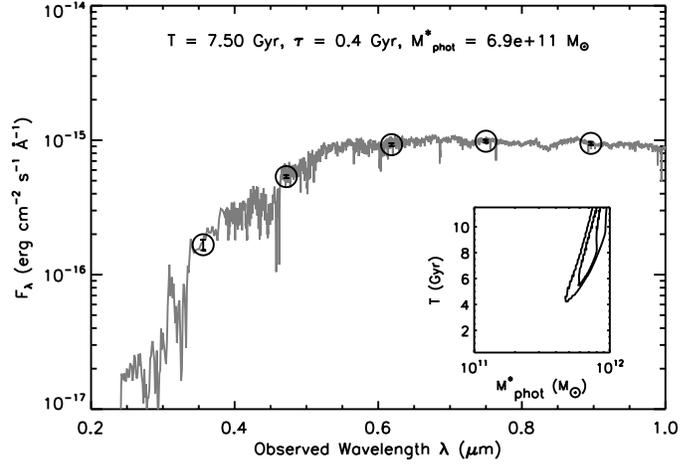}
  \caption{SED and best-fit model of the lens galaxy SDSS J0912+0029 at $z=0.1642$. The circles with the error bars represent, from left to right, the observed total flux densities measured in the \emph{u}, \emph{g}, \emph{r}, \emph{i}, and \emph{z} SDSS passbands. On the top, the best-fit values of the age ($T$), the characteristic  time of the SFH ($\mathrm{\tau}$), and the mass ($M_{\mathrm{phot}}^{*}$) of the galaxy are given. On the bottom right, the inset shows the 68\% and 99\% confidence regions for $T$ and $M_{\mathrm{phot}}^{*}$.}
    \label{sdss9}
\end{figure}   

The flux of our galaxies has been measured in five different bands (see Table \ref{Infotable}), from 354 to 913 nm (\emph{ugriz} filters of the SDSS). In order to obtain unbiased galaxy colors, in the absence of color gradients, a de Vaucouleurs profile,
\begin{equation}
I(R)=I_{0}\,\mathrm{exp}\left(-7.67\left(\frac{R}{R_{e}}\right)^{\nicefrac{1}{4}}\right)\,
\label{eq:1}
\end{equation}
($R_{e}$ being the standard optical effective radius), is fitted in the \emph{r} band of each object and this model is then varied, only in amplitude, for the other bands (after convolution with the corresponding point spread function). The resulting magnitudes are called $\mathrm{modelMag}$ in the publicly available SDSS catalog and correspond to magnitudes measured through equivalent apertures in all bands.

We derive the total photometric stellar mass ($M_{\mathrm{phot}}^{*}$) by fitting the observed SED with a three-parameter grid of composite stellar population (CSP) models. We use Bruzual \& Charlot's (\cite{bru03}) templates at solar metallicity, assuming a Salpeter (\cite{sal}) time-independent initial mass function (IMF) and a delayed exponential star formation history (SFH) parametrized by a characteristic timescale ($\tau$). The other two free parameters are the age of the model ($T$), which is constrained by the age of the Universe at the galaxy redshift, and the stellar mass. The uncertainties on the best-fit parameters are estimated by projecting the joint probability density distribution onto the corresponding axes. As an example, Fig. \ref{sdss9} shows the SED of a galaxy in our sample and the best-fit model. The estimated stellar mass values are given in Table \ref{Masstable}. We emphasize that the accuracy and homogeneity of the SDSS photometry, particularly the lack of significant systematics due to calibration or aperture corrections, ensure accurate photometric mass estimates. 

The photometric stellar mass inside the Einstein radius, $M_{\mathrm{phot}}^{*}(\le R_{\mathrm{Ein}})$, of Table \ref{Masstable} is inferred by multiplying the total mass by an aperture factor ($f_{\mathrm{ap}}$):
\begin{equation}
M_{\mathrm{phot}}^{*}(\le R_{\mathrm{Ein}}) = f_{\mathrm{ap}}\,\times\,M_{\mathrm{phot}}^{*}\,.
\label{eq:5}
\end{equation} 
This last factor represents the fraction of light enclosed inside the Einstein radius, with respect to the total light of the galaxy parametrized by the de Vaucouleurs profile, and is explicitly given by the following expression:
\begin{equation}
f_{\mathrm{ap}} = \frac{\int_{0}^{R_{\mathrm{Ein}}}I(R)\,R\,dR}{\int_{0}^{\infty}I(R)\,R\,dR}\,.
\label{eq:2}
\end{equation}
The assumption implicit in using Eq. (\ref{eq:5}) is that the stellar mass is traced by the light distribution.

Finally, we note that stellar mass estimates depend on the adopted IMF and SFH, but not on the different stellar population models (e.g. Bruzual \& Charlot \cite{bru03} vs. Maraston \cite{mar05}). We will come back to these points in the next section.

%                                     Two column figure (place early!)
%______________________________________________ Gamma_1 (lg rho, lg e)

%
%______________________________________________________________

\section{Comments and conclusions}

\begin{figure}
  \centering
  \includegraphics[width=0.49\textwidth]{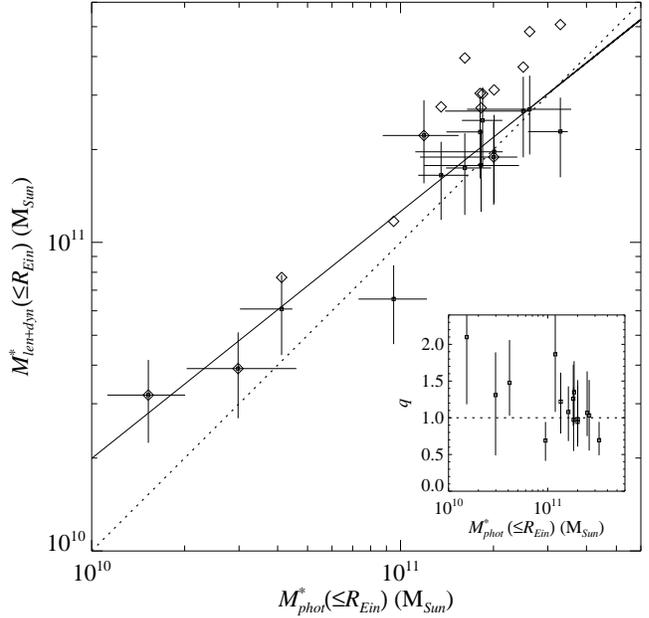}
  \caption{Comparison of the lensing+dynamical and photometric stellar masses (measured inside the Einstein radii) for the SLACS sample of gravitational lens early-type galaxies. The plot shows $M_{\mathrm{len+dyn}}^{*}(\le R_{\mathrm{Ein}})$ vs. $M_{\mathrm{phot}}^{*}(\le R_{\mathrm{Ein}})$, with the best fit correlation (solid) and the $M_{\mathrm{len+dyn}}^{*}(\le R_{\mathrm{Ein}})=M_{\mathrm{phot}}^{*}(\le R_{\mathrm{Ein}})$ (dotted) lines. The diamonds indicate upper limits for the lensing+dynamical stellar masses obtained by assuming that the measured masses enclosed inside the Einstein circles are only stellar. In the inset, we plot the ratio ($q$) between the lensing+dynamical and photometric stellar mass estimates as a function of $M_{\mathrm{phot}}^{*}(\le R_{\mathrm{Ein}})$.}
\label{mass_comp}
\end{figure}

In Fig. \ref{mass_comp}, we compare our photometric stellar mass estimates with the lensing+dynamics ones of Koopmans et al. (\cite{koo3}). This plot shows that the two different mass measurements are statistically correlated and consistent within the error bars. In particular, the value of the Pearson linear correlation coefficient is $\rho = 0.94$, and the best-fit correlation line yields:
\begin{equation}
M_{\mathrm{len+dyn}}^{*}(\le R_{\mathrm{Ein}})=10^{2.30\pm1.68}\times M_{\mathrm{phot}}^{*}(\le R_{\mathrm{Ein}})^{0.80\pm0.15}\,.
\label{eq:2}
\end{equation}
The median value of the ratio ($q$) between the lensing+dynamical and photometric stellar mass estimates is consistent with unity ($1.1\pm0.1$). The ratio $q$ does not show any correlation with galaxy colors, hence excluding a possible source of systematic errors in the photometric mass estimates. 

No significant difference in the relation between $M_{\mathrm{len+dyn}}^{*}(\le\,R_{\mathrm{Ein}})$ and $M_{\mathrm{phot}}^{*}(\le\,R_{\mathrm{Ein}})$ is observed deriving photometric stellar masses with Maraston's (\cite{mar05}) CSP models, as the two stellar population models differ remarkably in a near-IR regime at wavelengths longer than the ones probed by the SDSS filters. This result is in agreement with that reported by Rettura et al. (\cite{ret06}). By choosing Kroupa (\cite{kro01}) or Chabrier's (\cite{cha03}) IMFs, the photometric mass measurements are lowered in such a way that the slope of the best fit is consistent with that of Fig. \ref{mass_comp}, but the average ratio between the two mass estimates is considerably smaller than one. Moreover, we note that the effect of contamination caused by the lensed objects in the measurements of the fluxes of a lens galaxy is small. This is usually more relevant in the bluer filters, which are known to be less sensitive to photometric mass estimates, and the measurements of total magnitudes through de Vaucouleurs profile fitting should further reduce this source of uncertainty.

Values of $q$ slightly larger than one, like those observed and shown in the inset of Fig. \ref{mass_comp}, may be explained by possible underestimates of $M_{\mathrm{phot}}^{*}(\le R_{\mathrm{Ein}})$, which can be ascribed to two different phenomena occurring in the galaxies: dust extinction and metallicity values lower than the solar one. In detail, both effects tend to give lower IR fluxes, which then result in lower mass estimates. Nevertheless, several tests have supported the validity of the dust-free and solar metallicity model assumptions (e.g., see Rettura et al. \cite{ret06}).

Finally, despite a number of assumptions, we conclude that the good agreement between the two mass estimators, within their respective uncertainties, is a very reassuring result. This makes the presence of strong biases in one of the two methods very unlikely, and allows us to use either of them independently to measure reliably stellar masses. Although this study is based on a low redshift sample, we expect photometric mass estimates to be also accurate to high redshift, as long as the same optical/near-IR rest frame bands are covered.

\begin{acknowledgements}
      
\end{acknowledgements}
We thank G. Bertin for useful comments on this manuscript. We acknowledge the use of data from the accurate SDSS database. The SDSS Web Site is http://www.sdss.org/.

\begin{landscape}
\begin{table}
\centering
\caption{The relevant spectroscopic and photometric measurements of the SLACS sample.}
\label{Infotable}
\begin{tabular}{lcccccccc} 
\hline\hline             
SDSS Name & $z$ & $R_{e}$ & $R_{\mathrm{Ein}}$ & $u$ & $g$ & $r$ & $i$ & $z$ \\
 & & (\arcsec) & (\arcsec) & (mag) & (mag) & (mag) & (mag) & (mag) \\
\hline
J0037$-$0942 & 0.1955 & 2.38 $\pm$ 0.02 & 1.47 & 19.740 $\pm$ 0.127 & 18.038 $\pm$ 0.010 & 16.807 $\pm$ 0.006 & 16.339 $\pm$ 0.006 & 16.013 $\pm$ 0.016 \\
J0216$-$0813 & 0.3317 & 3.37 $\pm$ 0.22 & 1.15 & 21.034 $\pm$ 0.335 & 19.124 $\pm$ 0.024 & 17.455 $\pm$ 0.009 & 16.860 $\pm$ 0.009 & 16.593 $\pm$0.021 \\
J0737$+$3216 & 0.3223 & 3.26 $\pm$ 0.13 & 1.03 & 21.200 $\pm$ 0.266 & 19.400 $\pm$ 0.025 & 17.834 $\pm$ 0.010 & 17.214 $\pm$ 0.008 & 16.892 $\pm$ 0.020 \\
J0912$+$0029 & 0.1642 & 4.81 $\pm$ 0.02 & 1.61 & 19.287 $\pm$ 0.063 & 17.410 $\pm$ 0.007 & 16.228 $\pm$ 0.004 & 15.746 $\pm$ 0.004 & 15.399 $\pm$ 0.008 \\
J0956$+$5100 & 0.2405 & 2.60 $\pm$ 0.03 & 1.32 & 20.134 $\pm$ 0.136 & 18.475 $\pm$ 0.012 & 17.129 $\pm$ 0.007 & 16.632 $\pm$ 0.006 & 16.267 $\pm$ 0.013 \\
J0959$+$0410 & 0.1260 & 1.82 $\pm$ 0.05 & 1.00 & 20.363 $\pm$ 0.088 & 18.697 $\pm$ 0.012 & 17.639 $\pm$ 0.007 & 17.169 $\pm$ 0.006 & 16.783 $\pm$ 0.016 \\
J1250$+$0523 & 0.2318 & 1.77 $\pm$ 0.01 & 1.15 & 19.943 $\pm$ 0.084 & 18.500 $\pm$ 0.012 & 17.256 $\pm$ 0.007 & 16.732 $\pm$ 0.006 & 16.484 $\pm$ 0.014 \\
J1330$-$0148 & 0.0808 & 1.23 $\pm$ 0.01 & 0.85 & 20.060 $\pm$ 0.081 & 18.371 $\pm$ 0.009 & 17.442 $\pm$ 0.006 & 17.063 $\pm$ 0.007 & 16.742 $\pm$ 0.015 \\
J1402$+$6321 & 0.2046 & 3.14 $\pm$ 0.02 & 1.39 & 20.353 $\pm$ 0.142 & 18.294 $\pm$ 0.011 & 16.952 $\pm$ 0.006 & 16.444 $\pm$ 0.005 & 16.097 $\pm$ 0.011 \\
J1420$+$6019 & 0.0629 & 2.60 $\pm$ 0.10 & 1.04 & 18.156 $\pm$ 0.025 & 16.386 $\pm$ 0.004 & 15.541 $\pm$ 0.003 & 15.153 $\pm$ 0.003 & 14.889 $\pm$ 0.016 \\
J1627$-$0053 & 0.2076 & 2.14 $\pm$ 0.02 & 1.21 & 20.583 $\pm$ 0.190 & 18.588 $\pm$ 0.017 & 17.286 $\pm$ 0.008 & 16.805 $\pm$ 0.008 & 16.510 $\pm$ 0.017 \\
J1630$+$4520 & 0.2479 & 2.02 $\pm$ 0.02 & 1.81 & 20.554 $\pm$ 0.138 & 18.876 $\pm$ 0.015 & 17.396 $\pm$ 0.007 & 16.861 $\pm$ 0.007 & 16.561 $\pm$ 0.014 \\
J2300$+$0022 & 0.2285 & 1.80 $\pm$ 0.01 & 1.25 & 20.476 $\pm$ 0.190 & 19.007 $\pm$ 0.017 & 17.647 $\pm$ 0.009 & 17.126 $\pm$ 0.008 & 16.803 $\pm$ 0.022 \\
J2303$+$1422 & 0.1553 & 4.20 $\pm$ 0.04 & 1.64 & 19.427 $\pm$ 0.194 & 17.562 $\pm$ 0.012 & 16.385 $\pm$ 0.006 & 15.907 $\pm$ 0.006 & 15.605 $\pm$ 0.014 \\
J2321$-$0939 & 0.0819 & 4.47 $\pm$ 0.01 & 1.57 & 18.045 $\pm$ 0.037 & 16.145 $\pm$ 0.004 & 15.200 $\pm$ 0.003 & 14.772 $\pm$ 0.003 & 14.478 $\pm$ 0.006 \\
\hline
\end{tabular}
\begin{list}{}{}
\item[Notes --]Magnitudes are extinction-corrected modelMag (AB) from the SDSS.
\item[References --]Treu et al. (\cite{tre1}); Koopmans et al. (\cite{koo3}).
\end{list}
\end{table}

\begin{table}
\centering
\caption{The lensing+dynamical and photometric mass measurements of the SLACS sample.}
\label{Masstable}
\begin{tabular}{lcccccc} 
\hline\hline             
SDSS Name & $M_{\mathrm{len+dyn}}^{\mathrm{tot}}(\le R_{\mathrm{Ein}})$ & $f_{*}$ & $M_{\mathrm{phot}}^{*}$ & $f_{\mathrm{ap}}$ & $M_{\mathrm{len+dyn}}^{*}(\le R_{\mathrm{Ein}})$ & $M_{\mathrm{phot}}^{*}(\le R_{\mathrm{Ein}})$ \\
 & ($10^{10}$ M$_{\sun}$) & & ($10^{10}$ M$_{\sun}$) & & ($10^{10}$ M$_{\sun}$) & ($10^{10}$ M$_{\sun}$) \\
\hline
J0037$-$0942 & 27.3 & 0.65 $\pm$ 0.19 & $49^{+16}_{-17}$ & 0.37 & $18\pm5$ & $18^{+6}_{-6}$ \\
J0216$-$0813 & 48.2 & 0.56 $\pm$ 0.16 & $110^{+40}_{-41}$ &  0.24 & $27\pm8$ & $26^{+10}_{-10}$ \\
J0737$+$3216 & 31.2 & 0.63 $\pm$ 0.20 & $90^{+6}_{-40}$ & 0.22 & $20\pm6$ & $20^{+1}_{-9}$ \\
J0912$+$0029 & 39.6 & 0.44 $\pm$ 0.13 & $69^{+15}_{-9}$ & 0.23 & $17\pm5$ & $16^{+4}_{-2}$ \\
J0956$+$5100 & 37.0 & 0.72 $\pm$ 0.21 & $77^{+4}_{-34}$ & 0.32 & $27\pm8$ & $25^{+1}_{-11}$ \\
J0959$+$0410 & 7.7 & 0.79 $\pm$ 0.23 & $12^{+1}_{-3}$ & 0.34 & $6\pm2$ & $4^{+1}_{-1}$ \\
J1250$+$0523 & 18.9 & 1.04 $\pm$ 0.30 & $52^{+10}_{-22}$ & 0.39 & $19\pm6$ & $20^{+4}_{-8}$ \\
J1330$-$0148 & 3.2 & 1.05 $\pm$ 0.30 & $4^{+1}_{-1}$ & 0.40 & $3\pm1$ & $2^{+1}_{-1}$ \\
J1402$+$6321 & 30.3 & 0.82 $\pm$ 0.23 & $63^{+10}_{-9}$ & 0.29 & $25\pm7$ & $18^{+3}_{-3}$ \\
J1420$+$6019 & 3.9 & 1.08 $\pm$ 0.31 & $11^{+6}_{-4}$ & 0.27 & $4\pm1$ & $3^{+2}_{-1}$ \\
J1627$-$0053 & 22.2 & 1.04 $\pm$ 0.30 & $34^{+10}_{-9}$ & 0.35 & $22\pm7$ & $12^{+4}_{-3}$ \\
J1630$+$4520 & 50.8 & 0.45 $\pm$ 0.13 & $70^{+4}_{-15}$ & 0.47 & $23\pm7$ & $33^{+2}_{-7}$ \\
J2300$+$0022 & 30.4 & 0.75 $\pm$ 0.22 & $45^{+1}_{-10}$ & 0.40 & $23\pm7$ & $18^{+1}_{-4}$ \\
J2303$+$1422 & 27.5 & 0.60 $\pm$ 0.17 & $51^{+11}_{-8}$ & 0.27 & $17\pm5$ & $14^{+3}_{-2}$ \\
J2321$-$0939 & 11.7 & 0.56 $\pm$ 0.16 & $39^{+11}_{-9}$ & 0.24 & $7\pm2$ & $9^{+3}_{-2}$ \\
\hline
\end{tabular}
\begin{list}{}{}
\item[References --]Koopmans et al. (\cite{koo3}).
\end{list}
\end{table}
\end{landscape}

\end{document}